\documentclass{article}[10pt]
\usepackage{spconf,amssymb,amsmath,mathrsfs}
\usepackage{color,array,times}
\usepackage{epsfig,textcomp,epstopdf} 
\begin{document}

\title{Study of Efficient Robust Adaptive Beamforming Algorithms Based on Shrinkage Techniques}

\name{Hang Ruan * ~and Rodrigo C. de Lamare*$^\#$ \vspace{-0.6em} }

\address{ *Department of Electronics, The University of York, England, YO10 5BB\\
{ $^\#$CETUC, Pontifical Catholic University of Rio de Janeiro, Brazil}\\
 Emails: hr648@york.ac.uk,  delamare@cetuc.puc-rio.br \vspace{-0.85em}
 \sthanks{This work was supported in part by The
University of York}}

\maketitle

\begin{abstract}
This paper proposes low-complexity robust adaptive beamforming (RAB) techniques based on shrinkage methods. We firstly briefly review a Low-Complexity Shrinkage-Based Mismatch Estimation (LOCSME) batch algorithm to estimate the desired signal steering vector mismatch, in which the interference-plus-noise covariance (INC) matrix is also estimated with a recursive matrix shrinkage method. Then we develop low complexity adaptive robust version of the conjugate gradient (CG) algorithm to both estimate the steering vector mismatch and update the beamforming weights. A computational complexity study of the proposed and existing algorithms is carried out. Simulations are conducted in local scattering scenarios and comparisons to existing RAB techniques are provided.
\end{abstract}


\section{introduction}

Sensor array signal processing techniques and their applications to wireless
communications, sensor networks and radar have been widely investigated in
recent years. Adaptive beamforming is one of the most important topics in
sensor array signal processing which has applications in many fields. However,
adaptive beamformers may suffer performance degradation due to small sample
data size or the presence of the desired signal in the training data. In
practical environments, desired signal steering vector mismatch problems like
signal pointing errors \cite{r16}, imprecise knowledge of the antenna array,
look-direction mismatch or local scattering may even lead to more significant
performance loss \cite{r4}.

\subsection{{Prior and Related Work}}

In order to address these problems, robust adaptive beamforming (RAB)
techniques have been developed in recent years. Popular approaches include
worst-case optimization \cite{r4}, diagonal loading \cite{r5,r6}, and
eigen-decomposition \cite{r15,r16}. However, general RAB designs have some
limitations such as their ad hoc nature, high probability of subspace swap at
low SNR and high computational cost \cite{r7}.

Further recent works have looked at approaches based on combined estimation
procedures for both the steering vector mismatch and interference-plus-noise
covariance (INC) matrix to improve RAB performance. The worst-case optimization
methods in \cite{r4} solve an online semi-definite programming (SDP) while
using a matrix inversion to estimate the INC matrix. The method in \cite{r10}
estimates the steering vector mismatch by solving an online Sequential
Quadratic Program (SQP) \cite{r8}, while estimating the INC matrix using a
shrinkage method \cite{r10}. Another similar method which jointly estimates the
steering vector using SQP and the INC matrix using a covariance reconstruction
method \cite{r11} has outstanding performance compared to other RAB techniques.
However, their main disadvantages include the high computational cost
associated with online optimization programming, the matrix inversion or
reconstruction process, and slow convergence. In \cite{r14} and \cite{r19} we
have introduced a Low-Complexity Shrinkage-Based Mismatch Estimation (LOCSME)
method and a reduced-cost adaptive version \cite{r19} for robust beamforming,
which estimate the steering vector mismatch by exploiting the cross-correlation
vector between the sensor array data and the beamformer output.

\subsection{{Contributions}}

In this work, we develop an adaptive version of the LOCSME technique in
\cite{r14} based on a conjugate gradient (CG) adaptive algorithm, resulting in
the proposed LOCSME-CG algorithm. Different from the approach of LOCSME-SG, the
LOCSME-CG algorithm not only updates the beamforming weights using a subspace
approach \cite{scharf}-\cite{saalt}, but can also estimates the mismatched
steering vector, which sequentially performs the estimation of the mismatched
vector by LOCSME in every snapshot for large systems
\cite{mmimo}-\cite{BDVP},\cite{delamare_mber}-\cite{mbdf}. An analysis shows
that LOCSME-CG requires lower complexity than the original LOCSME and has
comparable complexity to LOCSME-SG. Simulations also show an excellent
performance which benefits from the precise estimation of the steering vector.

The paper is organized as follows. The system model and problem statement are
described in Section II. A review of the LOCSME method is provided in Section
III whereas Section IV presents the proposed LOCSME-CG algorithm. Section V
presents the simulation results. Section VI gives the conclusion.

\section{System Model and Problem Statement}

Consider a linear antenna array of $M$ sensors and $K$ narrowband signals which impinge on the array. The data received at the $i$th snapshot can be modeled as
\begin{equation}
{\bf x}(i)={\bf A}({\boldsymbol \theta}){\bf s}(i)+{\bf n}(i), \label{eq1}
\end{equation}
where ${\bf s}(i) \in {\mathbb C}^{K \times 1}$ are uncorrelated source signals,
${\boldsymbol\theta}=[{\theta}_1,\dotsb,{\theta}_K]^T \in {\mathbb
R}^K$ is a vector containing the directions of arrival (DoAs), ${\bf
A}({\boldsymbol \theta})=[{\bf a}({\theta}_1 )+{\bf e}, \dotsb, {\bf
a}({\theta}_K)] \in {\mathbb C}^{M \times K}$ is the matrix which
contains the steering vector for each DoA and ${\bf e}$ is the
steering vector mismatch of the desired signal, ${\bf n}(i) \in
{\mathbb C}^{M \times 1}$ is assumed to be complex Gaussian noise
with zero mean and variance ${\sigma}^2_n$. The beamformer output is
\begin{equation}
y(i)={\bf w}^H{\bf x}(i), \label{eq2}
\end{equation}
where ${\bf w}=[w_1,\dotsb,w_M]^T \in {\mathbb C}^{M\times1}$ is the
beamformer weight vector, where $({\cdot})^H$ denotes the Hermitian
transpose. The optimum beamformer is computed by maximizing the
SINR given by
\begin{equation}
SINR=\frac{{\sigma}^2_1{\lvert{\bf w}^H{\bf a}\rvert}^2}{{\bf w}^H{\bf R}_{i+n}{\bf w}}. \label{eq3}
\end{equation}
where ${\sigma}^2_1$ is the desired signal power, ${\bf R}_{i+n}$ is the INC matrix. Assuming that the steering vector ${\bf a}$ is known precisely (${\bf a}={\bf a}({\theta}_1 )$), then problem \eqref{eq3} can be cast as an optimization problem
\begin{equation}
\begin{aligned}
& \underset{\bf w} {\text{minimize}}
&& {\bf w}^H{\bf R}_{i+n}{\bf w} \\
& \text{subject to} && {\bf w}^H{\bf a}=1, \label{eq4}
\end{aligned}
\end{equation}
which is known as the MVDR beamformer or Capon beamformer \cite{r1}. The optimum weight vector is given by ${\bf w}_{opt}=\frac{{{\bf R}^{-1}_{i+n}}{\bf a}}{{\bf a}^H{{\bf R}^{-1}_{i+n}}{\bf a}}.$ Since ${\bf R}_{i+n}$ is usually unknown in practice, it can be estimated by the sample covariance matrix (SCM) of the received data as
\begin{equation}
\hat{\bf R}(i)=\frac{1}{i}\sum\limits_{k=1}^i{\bf x}(k){{\bf x}^H}(k), \label{eq5}
\end{equation}
which results in the Sample Matrix Inversion (SMI) beamformer ${\bf
w}_{SMI}=\frac{\hat{\bf R}^{-1}{\bf a}}{{\bf a}^H\hat{\bf
R}^{-1}{\bf a}}$. However, the SMI beamformer requires a large
number of snapshots to converge and is sensitive to steering vector
mismatches \cite{r10,r11}. The problem we are interested in solving
is how to design low-complexity robust adaptive beamforming algorithms that
can preserve the SINR performance in the presence of uncertainties
in the steering vector of a desired signal.

\section{LOCSME Robust Beamforming Algorithm}

The basic idea of LOCSME \cite{r14} is to obtain a precise estimate of the desired
signal steering vector by exploiting cross-correlation vector between
the beamformer output and the array observation data and then computing the beamforming weights.

\subsection{{Steering Vector Estimation}}

The cross-correlation between the array observation data and the
beamformer output can be expressed as ${\bf d}=E\lbrace{\bf
x}y^*\rbrace$. With assumptions that ${\lvert{{\bf a}_m{\bf
w}}\rvert}\ll{\lvert{{\bf a}_1{\bf w}}\rvert}$ for $m=2,\dotsb,K$
and that the signal sources and that the system noise have zero mean
while the desired signal is independent from the interferers and the
noise, ${\bf d}$ can be rewritten as ${\bf
d}=E\lbrace{{{{\sigma}_1}^2{\bf a}_1^H{\bf w}{\bf a}_1}+{\bf n}{\bf
n}^H{\bf w}}\rbrace$. By projecting ${\bf d}$ onto a predefined
subspace \cite{r9}, which collects all possible information from the
desired signal, the unwanted part of ${\bf d}$ can be eliminated.
LOCSME also exploits prior knowledge which amounts to choosing an angular sector in which
the desired signal is located, say
$[{\theta}_1-{\theta}_e,{\theta}_1+{\theta}_e]$. The subspace
projection matrix ${\bf P}$ is given by
\begin{equation}
{\bf P}=[{\bf c}_1,{\bf c}_2,\dotsb,{\bf c}_p][{\bf c}_1,{\bf c}_2,\dotsb,{\bf c}_p]^H, \label{eq6}
\end{equation}
where ${\bf c}_1,\dotsb,{\bf c}_p$ are the $p$ principal eigenvectors of the matrix ${\bf C}$, which is defined by \cite{r8}
\begin{equation}
{\bf C}=\int\limits_{{\theta}_1-{\theta}_e}^{{\theta}_1+{\theta}_e}{\bf a}({\theta}){\bf a}^H({\theta})d{\theta}. \label{eq7}
\end{equation}
In order to achieve a better estimation of the steering vector, an extension of the oracle approximating shrinkage (OAS) (\cite{r12}) technique is employed to obtain a more accurate estimate of the vector ${\bf d}$. Let us define the sample correlation vector (SCV) in snapshot $i$ as
\begin{equation}
\hat{\bf l}(i)=\frac{1}{i}\sum\limits_{k=1}^i{\bf x}(k)y^*(k), \label{eq8}
\end{equation}
and its mean value as
\begin{equation}
\hat\nu(i)=\sum\hat{\bf l}(i)/M. \label{eq9}
\end{equation}
Then we aim to shrink the SCV towards its mean value $\hat\nu(i)$, which yields
\begin{equation}
\hat{\bf d}(i)=\hat\rho(i)\hat\nu(i)+(1-\hat\rho(i))\hat{\bf l}(i), \label{eq10}
\end{equation}
where $\hat\rho(i)$ represents the shrinkage cofficient ($\hat\rho(i) \in (0,1)$). To find out the optimum $\hat\rho(i)$, we minimize the mean square error (MSE) of $E[{\lVert\hat{\bf d}(i)-\hat{\bf d}(i-1)\rVert}^2]$, which leads to
\begin{multline}
\hat{\rho}(i) = \\
\frac{(1-\frac{2}{M})\hat{\bf d}^H(i-1)\hat{\bf l}(i-1) + \sum\hat{\bf d}(i-1)\sum^*\hat{\bf
d}(i-1)}{(i-\frac{2}{M})\hat{\bf d}^H(i-1)\hat{\bf l}(i-1)+(1-\frac{i}{M})\sum\hat{\bf d}(i-1)\sum^*\hat{\bf d}(i-1)}. \label{eq11}
\end{multline}
Once the correlation vector $\hat{\bf d}$ is obtained, the steering vector is estimated by
\begin{equation}
{\hat{\bf a}_1}(i)=\frac{{\bf P}\hat{\bf d}(i)}{{\lVert{{\bf P}\hat{\bf d}(i)}\rVert}_2}. \label{eq12}
\end{equation}

\subsection{{Signal Power Estimation and Beamforming Weights}}

Following the description in \cite{r14}, the desired signal power ${\sigma}^2_1$ is estimated by
\begin{equation}
\hat{\sigma}^2_1(i)=\frac{|{{\hat{\bf a}_1}^H}(i){\bf x}(i)|^2-|{{\hat{\bf a}_1}^H}(i){\hat{\bf a}_1}(i)|{\sigma}^2_n}{|{{\hat{\bf a}_1}^H}(i){\hat{\bf a}_1}(i)|^2}, \label{eq17}
\end{equation}
which has a linear complexity ${\mathcal O}(M)$.

Once the steering vector and power of the desired signal are obtained, the INC matrix is also estimated by a matrix shrinkage method \cite{r14} and the weight vector is computed by
\begin{equation}
\hat{\bf w}(i)=\frac{{\hat{\bf R}^{-1}_{i+n}}(i)\hat{\bf a}_1(i)}{{\hat{\bf a}_1}^H(i){\hat{\bf R}^{-1}_{i+n}}(i)\hat{\bf a}_1(i)}, \label{eq21}
\end{equation}
which has a computationally costly matrix inversion ${\hat{\bf R}^{-1}_{i+n}}(i)$.

\section{Proposed LOCSME-CG Algorithm}

In this section, we develop a CG adaptive strategy based on LOCSME. We employ the same recursions as in LOCSME to estimate the steering vector and the desired signal power, whereas the estimation procedure of the beamforming weights is different. In order to avoid costly inner recursions, we let only one iteration be performed per snapshot\cite{r17}. Here we denote the CG-based weights and steering vector updated by snapshots as
\begin{equation}
\hat{\bf a}_1(i)=\hat{\bf a}_1(i-1)+{\alpha}_{\hat{\bf a}_1}(i){\bf p}_{\hat{\bf a}_1}(i), \label{eq47}
\end{equation}
\begin{equation}
{\bf v}(i)={\bf v}(i-1)+{\alpha}_{\bf v}(i){\bf p}_{\bf v}(i). \label{eq48}
\end{equation}
As can be seen, the subscripts of all the quantities for inner iterations are eliminated. Then, we employ the degenerated scheme to ensure ${\alpha}_{\hat{\bf a}_1}(i)$ and ${\alpha}_{\bf v}(i)$ satisfy the convergence bound \cite{r17} given by
\begin{equation}
0\leq{\bf p}^H_{\hat{\bf a}_1}(i){\bf g}_{\hat{\bf a}_1}(i)\leq0.5{\bf p}^H_{\hat{\bf a}_1}(i){\bf g}_{\hat{\bf a}_1}(i-1), \label{eq49}
\end{equation}
\begin{equation}
0\leq{\bf p}^H_{\bf v}(i){\bf g}_{\bf v}(i)\leq0.5{\bf p}^H_{\bf v}(i){\bf g}_{\bf v}(i-1). \label{eq50}
\end{equation}
Instead of updating the negative gradient vectors ${\bf g}_{\hat{\bf a}_1}(i)$ and ${\bf g}_{\bf v}(i)$ in iterations, now we utilize the forgetting factor to re-express them in one snapshot as
\begin{multline}
{\bf g}_{\hat{\bf a}_1}(i)=(1-\lambda){\bf v}(i)+\lambda{\bf g}_{\hat{\bf a}_1}(i-1) \\
+\hat{\sigma}^2_1(i){\alpha}_{\hat{\bf a}_1}(i){\bf v}(i){\bf v}^H(i){\bf p}_{\hat{\bf a}_1}(i)-{\bf x}(i){\bf x}^H(i)\hat{\bf a}_1(i), \label{eq51}
\end{multline}
\begin{multline}
{\bf g}_{\bf v}(i)=(1-\lambda)\hat{\bf a}_1(i)+\lambda{\bf g}_{\bf v}(i-1)-{\alpha}_{\bf v}(i)(\hat{\bf R}(i) \\
-\hat{\sigma}^2_1(i)\hat{\bf a}_1(i)\hat{\bf a}^H_1(i)){\bf p}_{\bf v}(i)-{\bf x}(i){\bf x}^H(i){\bf v}(i-1). \label{eq52}
\end{multline}
Pre-multiplying \eqref{eq51} and \eqref{eq52} by ${\bf p}^H_{\hat{\bf a}_1}(i)$ and ${\bf p}^H_{\bf v}(i)$, respectively, and taking expectations we obtain
\begin{multline}
E[{\bf p}^H_{\hat{\bf a}_1}(i){\bf g}_{\hat{\bf a}_1}(i)]=E[{\bf p}^H_{\hat{\bf a}_1}(i)({\bf v}(i)-{\bf x}(i){\bf x}^H(i)\hat{\bf a}_1)(i)] \\
+{\lambda}E[{\bf p}^H_{\hat{\bf a}_1}(i){\bf g}_{\hat{\bf a}_1}(i-1)]-{\lambda}E[{\bf p}^H_{\hat{\bf a}_1}(i){\bf v}(i)] \\
+E[{\alpha}_{\hat{\bf a}_1}(i){\bf p}^H_{\hat{\bf a}_1}(i)\hat{\sigma}^2_1(i){\bf v}(i){\bf v}^H(i){\bf p}_{\hat{\bf a}_1}(i)], \label{eq53}
\end{multline}
\begin{multline}
E[{\bf p}^H_{\bf v}(i){\bf g}_{\bf v}(i)]={\lambda}E[{\bf p}^H_{\bf v}(i){\bf g}_{\bf v}(i-1)]-{\lambda}E[{\bf p}^H_{\bf v}(i)\hat{\bf a}_1(i)] \\
-E[{\alpha}_{\bf v}(i){\bf p}^H_{\bf v}(i)(\hat{\bf R}(i)-\hat{\sigma}^2_1(i)\hat{\bf a}_1(i)\hat{\bf a}^H_1(i)){\bf p}_{\bf v}(i)], \label{eq54}
\end{multline}
where in \eqref{eq54} we have $E[\hat{\bf R}(i){\bf v}(i-1)]=E[\hat{\bf a}_1(i)]$. After substituting \eqref{eq54} back into \eqref{eq50} we obtain the bounds for ${\alpha}_{\bf v}(i)$ as follows
\begin{multline}
\frac{(\lambda-0.5)E[{\bf p}^H_{\bf v}(i){\bf g}_{\bf v}(i-1)]-{\lambda}E[{\bf p}^H_{\bf v}(i)\hat{\bf a}_1(i)]}{E[{\bf p}^H_{\bf v}(i)(\hat{\bf R}(i)-\hat{\sigma}^2_1(i)\hat{\bf a}_1(i)\hat{\bf a}^H_1(i)){\bf p}_{\bf v}(i)]}{\leq}E[{\alpha}_{\bf v}(i)] \\
{\leq}\frac{{\lambda}E[{\bf p}^H_{\bf v}(i){\bf g}_{\bf v}(i-1)]-{\lambda}E[{\bf p}^H_{\bf v}(i)\hat{\bf a}_1(i)]}{E[{\bf p}^H_{\bf v}(i)(\hat{\bf R}(i)-\hat{\sigma}^2_1(i)\hat{\bf a}_1(i)\hat{\bf a}^H_1(i)){\bf p}_{\bf v}(i)]}. \label{eq55}
\end{multline}
Then we can introduce a constant parameter ${\eta}_{\bf v} \in [0,0.5]$ to restrict ${\alpha}_{\bf v}(i)$ within the bounds in \eqref{eq55} as
\begin{multline}
{\alpha}_{\bf v}(i)= \\
\frac{\lambda({\bf p}^H_{\bf v}(i){\bf g}_{\bf v}(i-1)-{\bf p}^H_{\bf v}(i)\hat{\bf a}_1(i))-{\eta}_{\bf v}{\bf p}^H_{\bf v}(i){\bf g}_{\bf v}(i-1)}{{\bf p}^H_{\bf v}(i)(\hat{\bf R}(i)-\hat{\sigma}^2_1(i)\hat{\bf a}_1(i)\hat{\bf a}^H_1(i)){\bf p}_{\bf v}(i)}. \label{eq56}
\end{multline}
Similarly, we can also obtain the bounds for ${\alpha}_{\hat{\bf a}_1}(i)$. For simplicity let us define $E[{\bf p}^H_{\hat{\bf a}_1}(i){\bf g}_{\hat{\bf a}_1}(i-1)]=A$, $E[{\bf p}^H_{\hat{\bf a}_1}(i){\bf v}(i)]=B$, $E[{\bf p}^H_{\hat{\bf a}_1}(i){\bf x}(i){\bf x}^H(i)\hat{\bf a}_1(i)]=C$ and $E[{\bf p}^H_{\hat{\bf a}_1}(i)\hat{\sigma}^2_1(i){\bf v}(i) \\ {\bf v}^H(i){\bf p}_{\hat{\bf a}_1}(i)]=D$. Substituting equation \eqref{eq53} into \eqref{eq49} gives
\begin{multline}
\frac{\lambda(B-A)-B+C}{D}{\leq}E[{\alpha}_{\hat{\bf a}_1}(i)] \\
{\leq}\frac{\lambda(B-A)-B+C+0.5A}{D}, \label{eq57}
\end{multline}
in which we can introduce another constant parameter ${\eta}_{\hat{\bf a}_1} \in [0,0.5]$ to restrict ${\alpha}_{\hat{\bf a}_1}(i)$ within the bounds in \eqref{eq57} as
\begin{equation}
E[{\alpha}_{\hat{\bf a}_1}(i)]=\frac{\lambda(B-A)-B+C+{\eta}_{\hat{\bf a}_1}A}{D}, \label{eq58}
\end{equation}
or
\begin{multline}
{\alpha}_{\hat{\bf a}_1}(i)=[\lambda({\bf p}^H_{\hat{\bf a}_1}(i){\bf v}(i)-{\bf p}^H_{\hat{\bf a}_1}(i){\bf g}_{\hat{\bf a}_1}(i-1))-{\bf p}^H_{\hat{\bf a}_1}(i){\bf v}(i) \\
+{\bf p}^H_{\hat{\bf a}_1}(i){\bf x}(i){\bf x}^H(i)\hat{\bf a}_1(i)+{\eta}_{\hat{\bf a}_1}{\bf p}^H_{\hat{\bf a}_1}(i){\bf g}_{\hat{\bf a}_1}(i-1)] \\
/[\hat{\sigma}^2_1(i){\bf p}^H_{\hat{\bf a}_1}(i){\bf v}(i){\bf v}^H(i){\bf p}_{\hat{\bf a}_1}(i)]. \label{eq59}
\end{multline}
Then we can update the direction vectors ${\bf p}_{\hat{\bf a}_1}(i)$ and ${\bf p}_{\bf v}(i)$ by
\begin{equation}
{\bf p}_{\hat{\bf a}_1}(i+1)={\bf g}_{\hat{\bf a}_1}(i)+{\beta}_{\hat{\bf a}_1}(i){\bf p}_{\hat{\bf a}_1}(i), \label{eq60}
\end{equation}
\begin{equation}
{\bf p}_{\bf v}(i+1)={\bf g}_{\bf v}(i)+{\beta}_{\bf v}(i){\bf p}_{\bf v}(i), \label{eq61}
\end{equation}
where ${\beta}_{\hat{\bf a}_1}(i)$ and ${\beta}_{\bf v}(i)$ are updated by
\begin{equation}
{\beta}_{\hat{\bf a}_1}(i)=\frac{[{\bf g}_{\hat{\bf a}_1}(i)-{\bf g}_{\hat{\bf a}_1}(i-1)]^H{\bf g}_{\hat{\bf a}_1}(i)}{{\bf g}^H_{\hat{\bf a}_1}(i-1){\bf g}_{\hat{\bf a}_1}(i-1)}, \label{eq62}
\end{equation}
\begin{equation}
{\beta}_{\bf v}(i)=\frac{[{\bf g}_{\bf v}(i)-{\bf g}_{\bf v}(i-1)]^H{\bf g}_{\bf v}(i)}{{\bf g}^H_{\bf v}(i-1){\bf g}_{\bf v}(i-1)}. \label{eq63}
\end{equation}
Finally we can update the beamforming weights by
\begin{equation}
{\bf w}(i)=\frac{{\bf v}(i)}{\hat{\bf a}^H_1(i){\bf v}(i)}, \label{eq64}
\end{equation}

To reproduce the proposed LOCSME-CG algorithm, equations \eqref{eq9}-\eqref{eq12},\eqref{eq17},\eqref{eq47},\eqref{eq48},\eqref{eq51},\eqref{eq52},\eqref{eq56},\eqref{eq59}-\eqref{eq64} are required. The forgetting factor $\lambda$ and constant $\eta$ for estimating $\alpha(i)$ need to be adjusted to give its best performance.

A complexity analysis in terms of flops (total number of additions and multiplications) required by the proposed LOCSME-CG algorithm and the existing ones is compared in Table \ref{table1}. The proposed LOCSME-CG algorithm avoids costly matrix inversion and multiplication procedures, which are unavoidable in the existing RAB algorithms. The LCWC algorithm of \cite{r15} requires $N$ inner iterations per snapshot, which significantly varies in different snapshots (here we take $N=50$ as an averagely evaluated value). It is clear that LOCSME-CG has lower complexity in terms of the number of sensors $M$, dominated by ${\mathcal O}(M^2)$, resulting in great advantages when $M$ is large.

\begin{table}
\small
\begin{center}
\caption{Complexity Comparison}
\begin{tabular}{|c|c|}
\hline
RAB Algorithms & Flops \\
\hline
LOCSME \cite{r14} & $4M^3+3M^2+20M$ \\
\hline
LOCSME-SG & $15M^2+30M$ \\
\hline
Algorithm of \cite{r10} & $M^{3.5}+7M^3+5M^2+3M$ \\
\hline
LOCME \cite{r9} & $2M^3+4M^2+5M$ \\
\hline
LCWC \cite{r15} & $100M^2+350M$ \\
\hline
LOCSME-CG & $13M^2+77M$ \\
\hline
\end{tabular} \label{table1}
\end{center}
\end{table}


\section{Simulation Results}

The simulations are carried out under both coherent and incoherent local scattering mismatch \cite{r6} scenarios. A uniform linear array (ULA) of $M=12$ omnidirectional sensors with half wavelength spacing is considered. $100$ repetitions are executed to obtain each point of the curves and a maximum of $i=300$ snapshots are observed. The desired signal is assumed to arrive at ${\theta}_1=10^\circ$ while there are other two interferers impinging on the antenna array from directions ${\theta}_2=30^\circ$ and ${\theta}_3=50^\circ$. The signal-to-interference ratio (SIR) is fixed at $0$dB. For our proposed algorithm, the angular sector in which the desired signal is assumed to be located is chosen as $[{\theta}_1-5^\circ,{\theta}_1+5^\circ]$ and the number of eigenvectors of the subspace projection matrix $p$ is selected manually with the help of simulations. The results focus on the beamformer output SINR performance versus the number of snapshots, or a variation of input SNR ($-10$dB to $30$dB).

\subsection{{Mismatch due to Coherent Local Scattering}}

The steering vector of the desired signal affected by a time-invariant coherent local scattering effect is modeled as
\begin{equation}
{\bf a}_1={\bf p}+\sum\limits_{k=1}^4{e^{j{\varphi}_k}}{\bf b}({\theta}_k), \label{eq83}
\end{equation}
where ${\bf p}$ corresponds to the direct path while ${\bf b}({\theta}_k)(k=1,2,3,4)$ corresponds to the scattered paths. The angles ${\theta}_k(k=1, 2, 3, 4)$ are randomly and independently drawn in each simulation run from a uniform generator with mean $10^\circ$ and standard deviation $2^\circ$. The angles ${\varphi}_k(k=1, 2, 3, 4)$ are independently and uniformly taken from the interval $[0,2\pi]$ in each simulation run. Notice that ${\theta}_k$ and ${\varphi}_k$ change from trials while remaining constant over snapshots.

\begin{figure}[!htb]
\begin{center}
\def\epsfsize#1#2{0.95\columnwidth}
\epsfbox{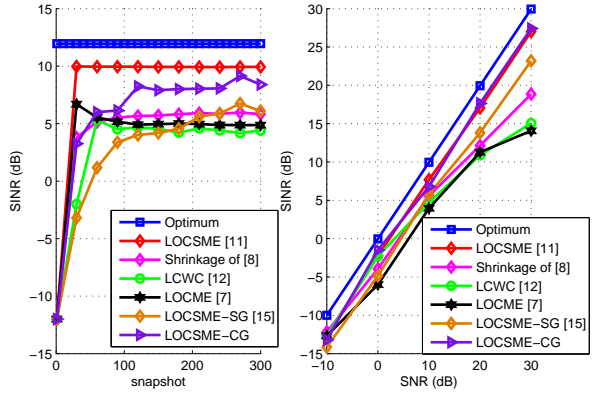}
\vspace{-0.2em}
\caption{coherent local scattering} \label{figure2}
\end{center}
\end{figure}

Fig. \ref{figure2} illustrate the performance comparisons of SINR versus snapshots and SINR versus SNR regarding the mentioned RAB algorithms in the last section under coherent scattering case. Specifically to obtain the SINR versus snapshots results, we select $\lambda=0.95$, $\eta=0.2$ for LOCSME-CG. However, selection of these parameters may vary according to different input SNR as in the SINR versus SNR results. LOCSME-CG outperforms the other algorithms and is very close to the standard LOCSME.

\subsection{{Mismatch due to Incoherent Local Scattering}}

In the incoherent local scattering case, the desired signal has a time-varying signature and the steering vector is modeled by
\begin{equation}
{\bf a}_1(i)=s_0(i){\bf p}+\sum\limits_{k=1}^4{s_k(i)}{\bf b}({\theta}_k), \label{eq84}
\end{equation}
where $s_k(i)(k=0, 1, 2, 3, 4)$ are i.i.d zero mean complex Gaussian random variables independently drawn from a random generator. The angles ${\theta}_k(k=0, 1, 2, 3, 4)$ are drawn independently in each simulation run from a uniform generator with mean $10^\circ$ and standard deviation $2^\circ$. This time, $s_k(i)$ changes both from run to run and from snapshot to snapshot.

\begin{figure}[!htb]
\begin{center}
\def\epsfsize#1#2{0.95\columnwidth}
\epsfbox{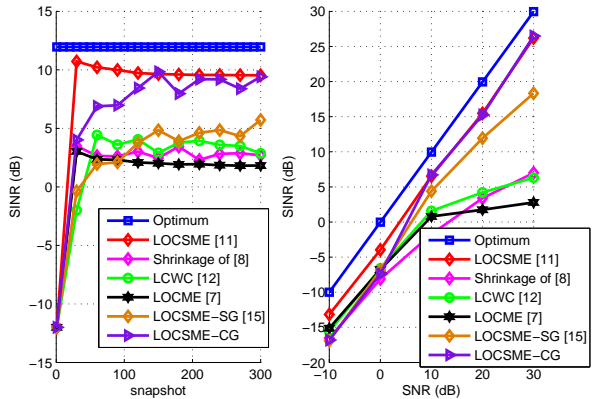}
\vspace{-0.2em}
\caption{incoherent local scattering} \label{figure3}
\end{center}
\end{figure}

Fig. \ref{figure3} illustrate the performance comparisons of SINR versus snapshots and SINR versus SNR regarding the mentioned RAB algorithms in the last section under incoherent scattering case. To obtain the SINR versus snapshots results, we select $\lambda=0.95$, $\eta=0.3$ for LOCSME-CG. However, we have optimized the parameters to give the best possible performance at different input SNRs.

Different from the coherent scattering results, all the algorithms have a certain level of performance degradation due to the effect of incoherent local scattering model, in which case we have the extra system dynamics with the time variation, contributing to more environmental uncertainties in the system. However, over a wide range of input SNR values, LOCSME-CG still outperforms the other RAB algorithms. One point that needs to be emphasized is, most of the existing RAB algorithms experience significant performance degradation when the input SNR is high (i.e. around or more than $20$dB), which is explained in \cite{r11} that the desired signal always presents in any kind of diagonal loading technique. However, LOCSME-CG improves the estimation accuracy, so that the high SNR degradation is successfully avoided as can be seen in Fig. \ref{figure2} and Fig. \ref{figure3}.

\section{Conclusion}

This work proposed a low-complexity adaptive RAB algorithm, LOCSME-CG, developed from the LOCSME RAB method. We have derived recursions for the weight vector update with low complexity. We also have enabled the estimation for the mismatch steering vector inside the CG recursions to enhance the robustness. Simulation results have shown that LOCSME-CG achieves excellent output SINR performance and is suitable for operation in high input SNR.


\begin{thebibliography}{10}
{\footnotesize
{\linespread{1.0}

\vspace{-1.0em}
\bibitem{r1}
H. L. Van Trees, {\it Optimum Array Processing}, New York: Wiley, 2002.

\vspace{-1.0em}
\bibitem{r4}
S. A. Vorobyov, A. B. Gershman and Z. Luo, ``Robust Adaptive Beamforming using Worst-Case Performance Optimization: A Solution to Signal Mismatch Problem," \emph{IEEE Trans. Sig. Proc.}, Vol. 51, No. 4, pp 313-324, Feb 2003.

\vspace{-1.0em}
\bibitem{r5}
J. Li, P. Stoica and Z. Wang, ``On Robust Capon Beamforming and Diagonal Loading," \emph{IEEE Trans. Sig. Proc.}, Vol. 57, No. 7, pp 1702-1715, July 2003.

\vspace{-1.0em}
\bibitem{r6}
D. Astely and B. Ottersten, ``The effects of Local Scattering on Direction of Arrival Estimation with Music," \emph{IEEE Trans. Sig. Proc.}, Vol. 47, No. 12, pp 3220-3234, Dec 1999.

\vspace{-1.0em}
\bibitem{r7}
A. Khabbazibasmenj, S. A. Vorobyov and A. Hassanien, ``Robust Adaptive Beamforming Based on Steering Vector Estimation with as Little as Possible Prior Information," \emph{IEEE Trans. Sig. Proc.}, Vol. 60, No. 6, pp 2974-2987, June 2012.

\vspace{-1.0em}
\bibitem{r8}
A. Hassanien, S. A. Vorobyov and K. M. Wong, ``Robust Adaptive Beamforming Using Sequential Quadratic Programming: An Iterative Solution to the Mismatch Problem," \emph{IEEE Sig. Proc. Letters.}, Vol. 15, pp 733-736, 2008.

\vspace{-1.0em}
\bibitem{r9}
L. Landau, R. de Lamare, M. Haardt, ``Robust Adaptive Beamforming Algorithms Using Low-Complexity Mismatch Estimation," \emph{Proc. IEEE Statistical Signal Processing Workshop}, 2011.

\vspace{-1.0em}
\bibitem{r10}
Y. Gu and A. Leshem, ``Robust Adaptive Beamforming Based on Jointly Estimating Covariance Matrix and Steering Vector," \emph{Proc. IEEE International Conference on Acoustics Speech and Signal Processing}, pp 2640-2643, 2011.

\vspace{-1.0em}
\bibitem{r11}
Y. Gu and A. Leshem, ``Robust Adaptive Beamforming Based on Interference Covariance Matrix Reconstruction and Steering Vector Estimation," \emph{IEEE Trans. Sig. Proc.}, Vol. 60, No. 7, July 2012.

\vspace{-1.0em}
\bibitem{r12}
Y. Chen, A. Wiesel and A. O. Hero III, ``Shrinkage Estimation of High Dimensional Covariance Matrices," \emph{Proc. IEEE International Conference on Acoustics Speech and Signal Processing}, pp 2937-2940, 2009.

\vspace{-1.0em}
\bibitem{r14}
H. Ruan and R. C. de Lamare, ``Robust Adaptive Beamforming Using a Low-Complexity Shrinkage-Based Mismatch Estimation Algorithm," \emph{IEEE Sig. Proc. Letters.}, Vol. 21, No. 1, pp 60-64, 2013.

\vspace{-1.0em}
\bibitem{r15}
A. Elnashar, ``Efficient implementation of robust adaptive beamforming based on worst-case performance optimization," IET Signal Process., Vol. 2, No. 4, pp. 381-393, Dec 2008.

\vspace{-1.0em}
\bibitem{r16}
J. Zhuang and A. Manikas, ``Interference cancellation beamforming robust to pointing errors," IET Signal Process., Vol. 7, No. 2, pp. 120-127, April 2013.

\vspace{-1.0em}
\bibitem{r17}
L. Wang and R. C. de Lamare, ``Constrained adaptive filtering algorithms based on conjugate gradient techniques for beamforming," IET Signal Process., Vol. 4, No. 6, pp. 686–697, Feb 2010.

\vspace{-1.0em}
\bibitem{r19}
H. Ruan and R. C. de Lamare, ``Low-Complexity Robust Adaptive Beamforming Based on Shrinkage and Cross-Correlation," \emph{19th International ITG Workshop on Smart Antennas}, pp 1-5, March 2015.



\bibitem{scharf}
L. L. Scharf and D. W. Tufts, ``Rank reduction for modeling stationary
signals," \textit{IEEE Transactions on Acoustics, Speech and Signal
Processing}, vol. ASSP-35, pp. 350-355, March 1987.



\bibitem{bar-ness} A. M. Haimovich
and Y. Bar-Ness, ``An eigenanalysis interference canceler," {\it IEEE Trans. on
Signal Processing}, vol. 39, pp. 76-84, Jan. 1991.

\bibitem{pados99} D. A. Pados and S. N. Batalama "Joint space-time
auxiliary vector filtering for DS/CDMA systems with antenna arrays" \textit{
IEEE Transactions on Communications}, vol. 47, no. 9, pp. 1406 - 1415, 1999.



\bibitem{reed98} J. S. Goldstein, I. S. Reed and L. L. Scharf
"A multistage representation of the Wiener filter based on orthogonal
projections" \textit{IEEE Transactions on Information Theory}, vol. 44, no. 7,
1998.

\bibitem{hua}
Y. Hua, M. Nikpour and P. Stoica, "Optimal reduced rank estimation and
filtering," IEEE Transactions on Signal Processing, pp. 457-469, Vol. 49, No.
3, March 2001.


\bibitem{goldstein}
M. L. Honig and J. S. Goldstein, ``Adaptive reduced-rank interference
suppression based on the multistage Wiener filter," \textit{ IEEE Transactions
on Communications}, vol. 50, no. 6, June 2002.

\bibitem{santos}
E. L. Santos and M. D. Zoltowski, ``On Low Rank MVDR Beamforming using the
Conjugate Gradient Algorithm", \textit{Proc. IEEE International Conference on
Acoustics, Speech and Signal Processing}, 2004.

\bibitem{qian}
Q. Haoli and S.N. Batalama, ``Data record-based criteria for the selection of
an auxiliary vector estimator of the MMSE/MVDR filter", \textit{IEEE
Transactions on Communications}, vol. 51, no. 10, Oct. 2003, pp. 1700 - 1708.

\bibitem{delamarespl07}
R. C. de Lamare and R. Sampaio-Neto, ``Reduced-Rank Adaptive Filtering Based on
Joint Iterative Optimization of Adaptive Filters", \textit{IEEE Signal
Processing Letters}, Vol. 14, no. 12, December 2007.

\bibitem{xutsa}
Z. Xu and M.K. Tsatsanis, ``Blind adaptive algorithms for minimum variance CDMA
receivers," \textit{IEEE Trans. Communications}, vol. 49, No. 1, January 2001.

\bibitem{delamaretsp}
R. C. de Lamare and R. Sampaio-Neto, ``Low-Complexity Variable Step-Size
Mechanisms for Stochastic Gradient Algorithms in Minimum Variance CDMA
Receivers", \textit{IEEE Trans. Signal Processing}, vol. 54, pp. 2302 - 2317,
June 2006.

\bibitem{kwak}
C. Xu, G. Feng and K. S. Kwak, ``A Modified Constrained Constant Modulus
Approach to Blind Adaptive Multiuser Detection," \textit{ IEEE Trans.
Communications}, vol. 49, No. 9, 2001.

\bibitem{xu&liu}
Z. Xu and P. Liu, ``Code-Constrained Blind Detection of CDMA Signals in
Multipath Channels," \textit{ IEEE Sig. Proc. Letters}, vol. 9, No. 12,
December 2002.

%
\bibitem{delamareccm}
R. C. de Lamare and R. Sampaio Neto, "Blind Adaptive Code-Constrained Constant
Modulus Algorithms for CDMA Interference Suppression in Multipath Channels",
\textit{ IEEE Communications Letters}, vol 9. no. 4, April, 2005.

\bibitem{wcccm}
L. Landau, R. C. de Lamare and M. Haardt, ``Robust adaptive beamforming
algorithms using the constrained constant modulus criterion," IET Signal
Processing, vol.8, no.5, pp.447-457, July 2014.

\bibitem{delamareelb}
R. C. de Lamare, ``Adaptive Reduced-Rank LCMV Beamforming Algorithms Based on
Joint Iterative Optimisation of Filters", \textit{Electronics Letters}, vol.
44, no. 9, 2008.


\bibitem{jidf}
R. C. de Lamare and R. Sampaio-Neto, ``Adaptive Reduced-Rank Processing Based
on Joint and Iterative Interpolation, Decimation and Filtering", \textit{IEEE
Transactions on Signal Processing}, vol. 57, no. 7, July 2009, pp. 2503 - 2514.

\bibitem{delamarecl}
R. C. de Lamare and Raimundo Sampaio-Neto, ``Reduced-rank Interference
Suppression for DS-CDMA based on Interpolated FIR Filters", \textit{IEEE
Communications Letters}, vol. 9, no. 3, March 2005.

\bibitem{delamaresp}
R. C. de Lamare and R. Sampaio-Neto, ``Adaptive Reduced-Rank MMSE Filtering
with Interpolated FIR Filters and Adaptive Interpolators", \textit{IEEE Signal
Processing Letters}, vol. 12, no. 3, March, 2005.

\bibitem{delamaretvt}
R. C. de Lamare and R. Sampaio-Neto, ``Adaptive Interference Suppression for
DS-CDMA Systems based on Interpolated FIR Filters with Adaptive Interpolators
in Multipath Channels", \textit{IEEE Trans. Vehicular Technology}, Vol. 56, no.
6, September 2007.

\bibitem{jioel}
R. C. de Lamare, ``Adaptive Reduced-Rank LCMV Beamforming Algorithms Based on
Joint Iterative Optimisation of Filters," Electronics Letters, 2008.


\bibitem{delamarespl07}
R. C. de Lamare and R. Sampaio-Neto, ``Reduced-rank adaptive filtering based on
joint iterative optimization of adaptive filters",  \textit{IEEE Signal
Process. Lett.}, vol. 14, no. 12, pp. 980-983, Dec. 2007.

\bibitem{delamare_ccmmswf}
R. C. de Lamare, M. Haardt, and R. Sampaio-Neto, ``Blind Adaptive Constrained
Reduced-Rank Parameter Estimation based on Constant Modulus Design for CDMA
Interference Suppression", \textit{IEEE Transactions on Signal Processing},
June 2008.

\bibitem{jidf_echo}
M. Yukawa, R. C. de Lamare and R. Sampaio-Neto, ``Efficient Acoustic Echo
Cancellation With Reduced-Rank Adaptive Filtering Based on Selective Decimation
and Adaptive Interpolation," IEEE Transactions on Audio, Speech, and Language
Processing, vol.16, no. 4, pp. 696-710, May 2008.

\bibitem{delamaretvt10}
R. C. de Lamare and R. Sampaio-Neto, ``Reduced-rank space-time adaptive
interference suppression with joint iterative least squares algorithms for
spread-spectrum systems," \textit{IEEE Trans. Vehi. Technol.}, vol. 59, no. 3,
pp. 1217-1228, Mar. 2010.

\bibitem{delamaretvt2011ST}
R. C. de Lamare and R. Sampaio-Neto, ``Adaptive reduced-rank equalization
algorithms based on alternating optimization design techniques for MIMO
systems," \textit{IEEE Trans. Vehi. Technol.}, vol. 60, no. 6, pp. 2482-2494,
Jul. 2011.


\bibitem{delamare10}
R. C. de Lamare, L. Wang, and R. Fa, ``Adaptive reduced-rank LCMV beamforming
algorithms based on joint iterative optimization of filters: Design and
analysis," Signal Processing, vol. 90, no. 2, pp. 640-652, Feb. 2010.

\bibitem{fa10}
R. Fa, R. C. de Lamare, and L. Wang, ``Reduced-Rank STAP Schemes for Airborne
Radar Based on Switched Joint Interpolation, Decimation and Filtering
Algorithm," \textit{IEEE Transactions on Signal Processing}, vol.58, no.8, Aug.
2010, pp.4182-4194.

\bibitem{lei09}
L. Wang and R. C. de Lamare, "Low-Complexity Adaptive Step Size Constrained
Constant Modulus SG Algorithms for Blind Adaptive Beamforming", \textit{Signal
Processing}, vol. 89, no. 12, December 2009, pp. 2503-2513.

\bibitem{ccmavf}
L. Wang and R. C. de Lamare, ``Adaptive Constrained Constant Modulus Algorithm
Based on Auxiliary Vector Filtering for Beamforming," IEEE Transactions on
Signal Processing, vol. 58, no. 10, pp. 5408-5413, Oct. 2010.


\bibitem{lei10}
L. Wang, R. C. de Lamare, M. Yukawa, "Adaptive Reduced-Rank Constrained
Constant Modulus Algorithms Based on Joint Iterative Optimization of Filters
for Beamforming," \textit{IEEE Transactions on Signal Processing}, vol.58,
no.6, June 2010, pp.2983-2997.

\bibitem{jio_ccm}
L. Wang, R. C. de Lamare and M. Yukawa, ``Adaptive reduced-rank constrained
constant modulus algorithms based on joint iterative optimization of filters
for beamforming", IEEE Transactions on Signal Processing, vol.58, no. 6, pp.
2983-2997, June 2010.

\bibitem{ccmavf}
L. Wang and R. C. de Lamare, ``Adaptive constrained constant modulus algorithm
based on auxiliary vector filtering for beamforming", IEEE Transactions on
Signal Processing, vol. 58, no. 10, pp. 5408-5413, October 2010.

\bibitem{stap_jio}
R. Fa and R. C. de Lamare, ``Reduced-Rank STAP Algorithms using Joint Iterative
Optimization of Filters," IEEE Transactions on Aerospace and Electronic
Systems, vol.47, no.3, pp.1668-1684, July 2011.

\bibitem{zhaocheng}
Z. Yang, R. C. de Lamare and X. Li, ``L1-Regularized STAP Algorithms With a
Generalized Sidelobe Canceler Architecture for Airborne Radar," IEEE
Transactions on Signal Processing, vol.60, no.2, pp.674-686, Feb. 2012.

\bibitem{zhaocheng2}
Z. Yang, R. C. de Lamare and X. Li, ``Sparsity-aware space–time adaptive
processing algorithms with L1-norm regularisation for airborne radar", IET
signal processing, vol. 6, no. 5, pp. 413-423, 2012.

\bibitem{arh_eusipco}
Neto, F.G.A.; Nascimento, V.H.; Zakharov, Y.V.; de Lamare, R.C., "Adaptive
re-weighting homotopy for sparse beamforming," in Signal Processing Conference
(EUSIPCO), 2014 Proceedings of the 22nd European , vol., no., pp.1287-1291, 1-5
Sept. 2014

\bibitem{arh_taes}
Almeida Neto, F.G.; de Lamare, R.C.; Nascimento, V.H.; Zakharov,
Y.V.,``Adaptive reweighting homotopy algorithms applied to beamforming," IEEE
Transactions on Aerospace and Electronic Systems, vol.51, no.3, pp.1902-1915,
July 2015.

\bibitem{dfjio}
L. Wang, R. C. de Lamare and M. Haardt, ``Direction finding algorithms based on
joint iterative subspace optimization," IEEE Transactions on Aerospace and
Electronic Systems, vol.50, no.4, pp.2541-2553, October 2014.

\bibitem{rdrab}
S. D. Somasundaram, N. H. Parsons, P. Li and R. C. de Lamare,
``Reduced-dimension robust capon beamforming using Krylov-subspace techniques,"
IEEE Transactions on Aerospace and Electronic Systems, vol.51, no.1,
pp.270-289, January 2015.

\bibitem{dcg_conf}
S. Xu and R.C de Lamare, , \textit{Distributed conjugate
gradient strategies for distributed estimation over sensor networks}, Sensor
Signal Processing for Defense SSPD, September 2012.


\bibitem{dcg}
S. Xu, R. C. de Lamare, H. V. Poor, ``Distributed Estimation Over Sensor
Networks Based on Distributed Conjugate Gradient Strategies", IET Signal
Processing, 2016 (to appear).

\bibitem{dce}
S. Xu, R. C. de Lamare and H. V. Poor, \textit{Distributed Compressed
Estimation Based on Compressive Sensing}, IEEE Signal Processing letters, vol.
22, no. 9, September 2014.

\bibitem{drr_conf}
S. Xu, R. C. de Lamare and H. V. Poor, ``Distributed reduced-rank estimation
based on joint iterative optimization in sensor networks," in Proceedings of
the 22nd European Signal Processing Conference (EUSIPCO), pp.2360-2364, 1-5,
Sept. 2014

\bibitem{dta_conf1}
S. Xu, R. C. de Lamare and H. V. Poor, ``Adaptive link selection strategies for
distributed estimation in diffusion wireless networks," in Proc. IEEE
International Conference onAcoustics, Speech and Signal Processing (ICASSP),  ,
vol., no., pp.5402-5405, 26-31 May 2013.

\bibitem{dta_conf2}
S. Xu, R. C. de Lamare and H. V. Poor, ``Dynamic topology adaptation for
distributed estimation in smart grids," in Computational Advances in
Multi-Sensor Adaptive Processing (CAMSAP), 2013 IEEE 5th International Workshop
on , vol., no., pp.420-423, 15-18 Dec. 2013.

\bibitem{dta_ls}
S. Xu, R. C. de Lamare and H. V. Poor, ``Adaptive Link Selection Algorithms for
Distributed Estimation", EURASIP Journal on Advances in Signal Processing,
2015.

\bibitem{song}
N. Song, R. C. de Lamare, M. Haardt, and M. Wolf, ``Adaptive Widely Linear
Reduced-Rank Interference Suppression based on the Multi-Stage Wiener Filter,"
IEEE Transactions on Signal Processing, vol. 60, no. 8, 2012.

\bibitem{wljio}
N. Song, W. U. Alokozai, R. C. de Lamare and M. Haardt, ``Adaptive Widely
Linear Reduced-Rank Beamforming Based on Joint Iterative Optimization,"  IEEE
Signal Processing Letters, vol.21, no.3, pp. 265-269, March 2014.

\bibitem{barc}
R.C. de Lamare, R. Sampaio-Neto and M. Haardt, "Blind Adaptive Constrained
Constant-Modulus Reduced-Rank Interference Suppression Algorithms Based on
Interpolation and Switched Decimation," \textit{IEEE Trans. on Signal
Processing},  vol.59, no.2, pp.681-695, Feb. 2011.

\bibitem{jiomber}
Y. Cai, R. C. de Lamare, ``Adaptive Linear Minimum BER Reduced-Rank
Interference Suppression Algorithms Based on Joint and Iterative Optimization
of Filters," IEEE Communications Letters, vol.17, no.4, pp.633-636, April 2013.

\bibitem{saalt}
R. C. de Lamare and R. Sampaio-Neto, ``Sparsity-Aware Adaptive Algorithms Based
on Alternating Optimization and Shrinkage," IEEE Signal Processing Letters,
vol.21, no.2, pp.225,229, Feb. 2014.

\bibitem{mmimo}
R. C. de Lamare, ``Massive MIMO Systems: Signal Processing Challenges and
Future Trends", Radio Science Bulletin, December 2013.

\bibitem{wence}
W. Zhang, H. Ren, C. Pan, M. Chen, R. C. de Lamare, B. Du and J. Dai,
``Large-Scale Antenna Systems With UL/DL Hardware Mismatch: Achievable Rates
Analysis and Calibration", IEEE Trans. Commun., vol.63, no.4, pp. 1216-1229,
April 2015.

\bibitem{Costa}
M. Costa, "Writing on dirty paper," \textit{IEEE Trans. Inform. Theory}, vol.
29, no. 3, pp. 439-441, May 1983.

\bibitem{delamare_ieeproc}
R. C. de Lamare and A. Alcaim, "Strategies to improve the performance of very
low bit rate speech coders and application to a 1.2 kb/s codec" IEE
Proceedings- Vision, image and signal processing, vol. 152, no. 1, February,
2005.

\bibitem{TDS_clarke}
P. Clarke and R. C. de Lamare, "Joint Transmit Diversity Optimization and Relay
Selection for Multi-Relay Cooperative MIMO Systems Using Discrete Stochastic
Algorithms," \emph{IEEE Communications Letters}, vol.15, no.10, pp.1035-1037,
October 2011.

\bibitem{TDS_2}
P. Clarke and R. C. de Lamare, "Transmit Diversity and Relay Selection
Algorithms for Multirelay Cooperative MIMO Systems" \emph{IEEE Transactions on
Vehicular Technology}, vol.61, no. 3, pp. 1084-1098, October 2011.

\bibitem{switch_int}
Y. Cai, R. C. de Lamare, and R. Fa, ``Switched Interleaving Techniques with
Limited Feedback for Interference Mitigation in DS-CDMA Systems," IEEE
Transactions on Communications, vol.59, no.7, pp.1946-1956, July 2011.

\bibitem{switch_mc}
Y. Cai, R. C. de Lamare, D. Le Ruyet, ``Transmit Processing Techniques Based on
Switched Interleaving and Limited Feedback for Interference Mitigation in
Multiantenna MC-CDMA Systems," IEEE Transactions on Vehicular Technology,
vol.60, no.4, pp.1559-1570, May 2011.

\bibitem{smce}
T. Wang, R. C. de Lamare, and P. D. Mitchell, ``Low-Complexity Set-Membership
Channel Estimation for Cooperative Wireless Sensor Networks," IEEE Transactions
on Vehicular Technology, vol.60, no.6, pp.2594-2607, July 2011.

\bibitem{TongW}
T. Wang, R. C. de Lamare and A. Schmeink, "Joint linear receiver design and
power allocation using alternating optimization algorithms for wireless sensor
networks," \textit{IEEE Trans. on Vehi. Tech.}, vol. 61, pp. 4129-4141, 2012.

\bibitem{jpais_iet}
R. C. de Lamare, ``Joint iterative power allocation and linear interference
suppression algorithms for cooperative DS-CDMA networks", IET Communications,
vol. 6, no. 13 , 2012, pp. 1930-1942.

\bibitem{TARMO}
T. Peng, R. C. de Lamare and A. Schmeink, ``Adaptive Distributed Space-Time
Coding Based on Adjustable Code Matrices for Cooperative MIMO Relaying
Systems'', \emph{IEEE Transactions on Communications}, vol. 61, no. 7, July
2013.

\bibitem{keke1}
K. Zu, R. C. de Lamare, ``Low-Complexity Lattice Reduction-Aided Regularized
Block Diagonalization for MU-MIMO Systems'', IEEE. Communications Letters, Vol.
16, No. 6, June 2012, pp. 925-928.

\bibitem{kekecl}
K. Zu, R. C. de Lamare, ``Low-Complexity Lattice Reduction-Aided Regularized
Block Diagonalization for MU-MIMO Systems'', IEEE. Communications Letters, Vol.
16, No. 6, June 2012.

\bibitem{keke2} K. Zu, R. C. de Lamare and M.
Haart, ``Generalized design of low-complexity block diagonalization type
precoding algorithms for multiuser MIMO systems", IEEE Trans. Communications,
2013.

\bibitem{Tomlinson}
M. Tomlinson, "New automatic equaliser employing modulo arithmetic,"
\textit{Electronic Letters}, vol. 7, Mar. 1971.

\bibitem{dopeg_cl} C. T. Healy and R. C. de Lamare,
``Decoder-optimised progressive edge growth algorithms for the design of LDPC
codes with low error floors",  \textit{IEEE Communications Letters}, vol. 16,
no. 6, June 2012, pp. 889-892.

\bibitem{peg_bf_iswcs}
A. G. D. Uchoa, C. T. Healy, R. C. de Lamare, R. D. Souza, ``LDPC codes based
on progressive edge growth techniques for block fading channels",
\textit{Proc. 8th International Symposium on Wireless Communication Systems
(ISWCS)}, 2011, pp. 392-396.

\bibitem{gqcpeg}
A. G. D. Uchoa, C. T. Healy, R. C. de Lamare, R. D. Souza, ``Generalised
Quasi-Cyclic LDPC codes based on progressive edge growth techniques for block
fading channels",  \textit{Proc. International Symposium Wireless Communication
Systems (ISWCS)}, 2012, pp. 974-978.

\bibitem{peg_bf_cl}
A. G. D. Uchoa, C. T. Healy, R. C. de Lamare, R. D. Souza, ``Design of LDPC
Codes Based on Progressive Edge Growth Techniques for Block Fading Channels",
\textit{IEEE Communications Letters}, vol. 15, no. 11, November 2011, pp.
1221-1223.

\bibitem{Harashima}
H. Harashima and H. Miyakawa, "Matched-transmission technique for channels with
intersymbol interference," \textit{IEEE Trans. Commun.}, vol. 20, Aug. 1972.

\bibitem{mbthpc}
K. Zu, R. C. de Lamare and M. Haardt, ``Multi-branch tomlinson-harashima
precoding for single-user MIMO systems," in Smart Antennas (WSA), 2012
International ITG Workshop on , vol., no., pp.36-40, 7-8 March 2012.

\bibitem{zuthp}
K. Zu, R. C. de Lamare and M. Haardt, ``Multi-Branch Tomlinson-Harashima
Precoding Design for MU-MIMO Systems: Theory and Algorithms," IEEE Transactions
on Communications, vol.62, no.3, pp.939,951, March 2014.


\bibitem{rmbthp}
L. Zhang, Y. Cai, R. C. de Lamare and M. Zhao,  ``Robust Multibranch
Tomlinson–Harashima Precoding Design in Amplify-and-Forward MIMO Relay
Systems," IEEE Transactions on Communications, vol.62, no.10, pp.3476,3490,
Oct. 2014.

\bibitem{Hochwald}
B. Hochwald, C. Peel and A. Swindlehurst, "A vector-perturbation technique for
near capacity multiantenna multiuser communication - Part II: Perturbation,"
\textit{IEEE Trans. Commun.}, vol. 53, no. 3, Mar. 2005.

\bibitem{BDVP}
C. B. Chae, S. Shim and R. W. Heath, "Block diagonalized vector perturbation
for multiuser MIMO systems," \textit{IEEE Trans. Wireless Commun.}, vol. 7, no.
11, pp. 4051 - 4057, Nov. 2008.

\bibitem{delamare_mber}
R. C. de Lamare, R. Sampaio-Neto, ``Adaptive MBER decision feedback multiuser
receivers in frequency selective fading channels", \textit{ IEEE Communications
Letters}, vol. 7, no. 2, Feb. 2003, pp. 73 - 75.

\bibitem{rontogiannis}
A. Rontogiannis, V. Kekatos, and K. Berberidis," A Square-Root Adaptive V-BLAST
Algorithm for Fast Time-Varying MIMO Channels," \textit{IEEE Signal Processing
Letters}, Vol. 13, No. 5, pp. 265-268, May 2006.

\bibitem{delamare_itic} R. C.
de Lamare, R. Sampaio-Neto, A. Hjorungnes, ``Joint iterative interference
cancellation and parameter estimation for CDMA systems", \textit{IEEE
Communications Letters}, vol. 11, no. 12, December 2007, pp. 916 - 918.

\bibitem{stspadf}
Y. Cai and R. C. de Lamare, "Adaptive Space-Time Decision Feedback Detectors
with Multiple Feedback Cancellation", \textit{IEEE Transactions on Vehicular
Technology}, vol. 58, no. 8,  October 2009, pp. 4129 - 4140.

\bibitem{choi}
J. W. Choi, A. C. Singer, J Lee, N. I. Cho, ``Improved linear soft-input
soft-output detection via soft feedback successive interference cancellation,"
\textit{IEEE Trans. Commun.}, vol.58, no.3, pp.986-996, March 2010.


\bibitem{stbcccm}
R. C. de Lamare and R. Sampaio-Neto, ``Blind adaptive MIMO receivers for
space-time block-coded DS-CDMA systems in multipath channels using the constant
modulus criterion," IEEE Transactions on Communications, vol.58, no.1,
pp.21-27, January 2010.


\bibitem{FL11}
R. Fa, R. C. de Lamare, ``Multi-Branch Successive Interference Cancellation for
MIMO Spatial Multiplexing Systems", \textit{ IET Communications}, vol. 5, no.
4, pp. 484 - 494, March 2011.

\bibitem{jio_mimo}
R.C. de Lamare and R. Sampaio-Neto, ``Adaptive reduced-rank equalization
algorithms based on alternating optimization design techniques for MIMO
systems," IEEE Trans. Veh. Technol., vol. 60, no. 6, pp. 2482-2494, July 2011.

\bibitem{peng_twc} P. Li, R. C. de Lamare and R. Fa, ``Multiple
Feedback Successive Interference Cancellation Detection for Multiuser MIMO
Systems," \textit{IEEE Transactions on Wireless Communications}, vol. 10, no.
8, pp. 2434 - 2439, August 2011.

\bibitem{spa}
R.C. de Lamare, R. Sampaio-Neto, ``Minimum mean-squared error iterative
successive parallel arbitrated decision feedback detectors for DS-CDMA
systems," IEEE Trans. Commun., vol. 56, no. 5, May 2008, pp. 778-789.

\bibitem{spa2}
R.C. de Lamare, R. Sampaio-Neto, ``Minimum mean-squared error iterative
successive parallel arbitrated decision feedback detectors for DS-CDMA
systems," IEEE Trans. Commun., vol. 56, no. 5, May 2008.

\bibitem{jio_mimo} R.C. de Lamare and R. Sampaio-Neto, ``Adaptive
reduced-rank equalization algorithms based on alternating optimization design
techniques for MIMO systems," IEEE Trans. Veh. Technol., vol. 60, no. 6, pp.
2482-2494, July 2011.

\bibitem{P.Li}
P. Li, R. C. de Lamare and J. Liu, ``Adaptive Decision Feedback Detection with
Parallel Interference Cancellation and Constellation Constraints for Multiuser
MIMO systems'', IET Communications, vol.7, 2012, pp. 538-547.

\bibitem{jingjing}
J. Liu, R. C. de Lamare, ``Low-Latency Reweighted Belief Propagation Decoding
for LDPC Codes," IEEE Communications Letters, vol. 16, no. 10, pp. 1660-1663,
October 2012.

\bibitem{did}
P. Li and R. C. de Lamare, Distributed Iterative Detection With Reduced Message
Passing for Networked MIMO Cellular Systems, IEEE Transactions on Vehicular
Technology, vol.63, no.6, pp. 2947-2954, July 2014.

\bibitem{bfidd}
A. G. D. Uchoa, C. T. Healy and R. C. de Lamare, ``Iterative Detection and
Decoding Algorithms For MIMO Systems in Block-Fading Channels Using LDPC
Codes," IEEE Transactions on Vehicular Technology, 2015.

\bibitem{mbdf} R.
C. de Lamare, "Adaptive and Iterative Multi-Branch MMSE Decision Feedback
Detection Algorithms for Multi-Antenna Systems", \emph{IEEE Trans. Wireless
Commun.}, vol. 14, no. 10, October 2013.

} }
\end{thebibliography}
\end{document}